\documentclass[aps,amsmath,amssymb]{revtex4}

\usepackage[rflt]{floatflt}
\usepackage{hyperref}
\usepackage{graphicx}
\begin{document}

\newcommand{\Det}{\,{\rm{Det}\,}}
\newcommand{\Tr}{\,{\rm{Tr}\,}}
\newcommand{\tr}{\,{\rm{tr}\,}}

\newcommand{\bomega}{\bar{\Omega}}
\newcommand{\wsf}{\omega_{\rm sf}}
\newcommand{\sign}{{\rm sign}}

\renewcommand{\Re}{{\rm Re}}
\renewcommand{\Im}{{\rm Im}}

\newcommand{\COMMENT}{\large\bf}
\newcommand{\FigWidth}{0.9\columnwidth}

\newcommand{\CC}{\mbox{\Bbb C}}
\newcommand{\NN}{\mbox{\Bbb N}}
\newcommand{\PP}{\mbox{\Bbb P}}        
\newcommand{\QQ}{\mbox{\Bbb Q}}
\newcommand{\RR}{\mbox{\Bbb R}}
\newcommand{\TT}{\mbox{\Bbb T}}
\newcommand{\XX}{\mbox{\Bbb X}}
\newcommand{\ZZ}{\mbox{$\mathbb Z$}}
\newcommand{\z}{\raise-1pt\hbox{$\mbox{\Bbbb Z}$}}

\def\beq{\begin{equation}}
\def\eeq{\end{equation}}
\def\p{\partial}
\def\G{\Gamma}

\def\MM{{\sf M}}
\def\DD{{\sf D}}
\def\Dc{\CC \setminus {\sf D}}
\def\Uc{{\sf U}_{{\rm ext}}}
\def\UU{{\sf U}}
\def\BB{{\sf B}}

\def\unitary{{\mathscr U}}
\def\hermitian{{\mathscr H}}
\def\normal{{\mathscr N}}
\def\complex{{\mathscr C}}

\def\kernel{K}
\def\lbracket{\left <}
\def\rbracket{\right >}
\def\vphcl{\varphi _{0}}
\def\rhocl{\rho _{0}}

\def\square{\hfill
{\vrule height6pt width6pt depth1pt} \break \vspace{.01cm}}

\title{Self-similarity in Laplacian Growth}

\author{Ar.~Abanov}
\affiliation{
            Department of Physics, MS 4242,
            Texas A\&M University,
            College Station, TX 77843-4242, USA
}
\author{M.~Mineev-Weinstein}
\affiliation{Los Alamos National Laboratory, MS-P365,
Los Alamos, NM 87545, USA
}
\author{A.~Zabrodin}
\affiliation{Institute of Biochemical Physics, Kosygina str. 4,
119991 Moscow, Russia; also at ITEP, Bol. Cheremushkinskaya str. 25,
117259 Moscow, Russia}

\date{\today}

\begin{abstract}
We consider Laplacian Growth of self-similar domains in different
geometries. Self-similarity determines the analytic structure of the
Schwarz function of the moving boundary. The knowledge of this
analytic structure allows us to derive the integral equation for the
conformal map. It is shown that solutions to the integral equation
obey also a second order differential equation which is the one
dimensional Schroedinger equation with the
$\mbox{sinh}^{-2}$-potential. The solutions, which are expressed
through the Gauss hypergeometric function, characterize the geometry
of self-similar patterns in a wedge. We also find the potential for
the Coulomb gas representation of the self-similar Laplacian growth
in a wedge and calculate the corresponding free energy.
\end{abstract}


\maketitle
\tableofcontents

\section{Introduction}

\begin{figure}
\includegraphics[width=\FigWidth]{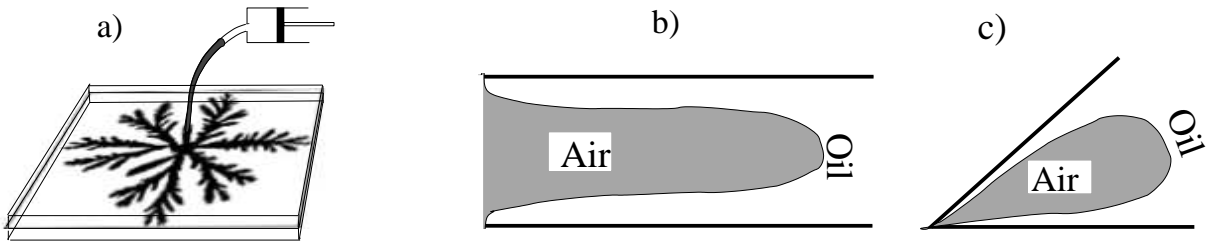}

\caption{\label{fig:LG} Laplacian Growth in a Hele-Shaw cell for the
radial {\it a}), channel {\it b}), and wedge {\it c}) geometries. }
\end{figure}

The Laplacian Growth (LG) phenomena has been studied for a long time
in various physical systems ranging from solidification to bacterial
colony growth \cite{Howison}. Displacement of viscous fluid by an
invicsid one (to be referred as {\it oil} and {\it air}) between two
closely spaced horizontal plates (the Hele-Shaw cell) is the
simplest process of this kind. The Laplacian Growth is defined here
as the dynamics of the oil/air interface (Fig. \ref{fig:LG}a).
For a review, see, e.g.,
\cite{RMP,book}.

The process is unstable and produces in a long time an asymptotical
shape which is either a finger \cite{ST,TRHC} (in channel and wedge
geometries) or a fractal with the same universal characteristics
\cite{HLS-DLA} as in diffusion-limited aggregation
\cite{WittenSanderPRL,PhTod} (in all geometries).

It has been recently discovered that the Laplacian Growth is
\begin{itemize}
\item
described by the 2D Toda integrable hierarchy \cite{MWZ},
\item
equivalent to the evolution of the support of eigenvalues in
certain ensembles of random matrices \cite{MWZ,KKMWZ},
\item
relevant to the 2D fermions in the magnetic field
\cite{ABWZ}, and
\item
equivalent to Whitham equations of the soliton theory \cite{KMWZ}.
\end{itemize}

The most well studied geometry is a rectangular Hele-Shaw cell
(Fig. \ref{fig:LG}b) where air enters the channel from the left
side. The boundary dynamics in this system was considered first by
Saffman and Taylor in 1958 \cite{ST}.  They observed a relatively
stable finger (known today as the Saffman-Taylor finger) formed
after a competition of many fingers generated by an interfacial
instability. They also found a continuous family of traveling wave
solutions (in the absence of surface tension) and posed a problem
of selection of a proper member of the family to describe the
observable finger. (See the development of the selection problem
in \cite{Asympt}-\cite{M98}). In the radial geometry (Fig.
\ref{fig:LG}a) an asymptotic shape is a fractal with nontrivial
universal self-similar features \cite{HLS-DLA}, which are still
not explained. Another geometry which has been studied both
theoretically and experimentally is the wedge geometry (Fig. \ref{fig:LG}c).
In this setting air enters the wedge filled with oil through the
wedge corner and pushes oil outward. A number of exact solutions
were found in this geometry by different methods
\cite{TRHC},\cite{Benamar}-\cite{MV}.
Some of them are
analogs of ``traveling wave'' solutions. They are also called
fingers.

The latter experimental setting is of the great interest for several
reasons. One of them is that traveling wave solutions in a wedge
geometry are the only known explicit solutions of LG described by
conformal maps (from the unit circle), which derivatives contain
non-integer singularities.  Moreover, they have a very important
property of being self-similar.

It is the latter property that we are going to study in detail in
this paper. We prove that the only smooth
self-similar solution in the
radial geometry is an ellipse. The family of self-similar
solutions with a singularity at the origin is much more
interesting. We show that the exact solutions in the wedge geometry
first obtained in \cite{Benamar} belong to this class.

To study self-similar solutions of general type,
we develop a new method which allows us to represent
the results for the wedge geometry in a more explicit
and transparent form. We will show that the
self-similarity completely determines the analytical structure of
the Schwarz function of the moving boundary.  Using the knowledge of
the analytical structure we will derive an integral equation for the
conformal map from the exterior of the unit circle to the domain
filled by oil.  We will establish then the equivalence of this
integral equation to the Schrodinger equation with a well-studied
(Poeschle-Teller) potential \cite{Flugge} which was derived for
LG in wedge earlier \cite{Benamar,Tu}.  We will also calculate the
free energy of the 2D Coulomb gas representation of LG for the case
of self-similar solutions in a wedge.

The paper is organized as follows. In Section \ref{sec:SF}, we
introduce the notion of the Schwarz function of a curve. In Section
\ref{sec:LGSF}, we formulate the Laplacian Growth in terms of the
Schwarz function of the moving boundary.  Section \ref{sec:SSSGR}
treats the property of self-similarity in terms of the
Schwarz-function. Sections \ref{sec:W} and \ref{sec:IndDiff} are
devoted to the self-similar solutions in the wedge. They contain
geometric interpretation, derivation of both integral and
differential equations for these solutions, proof of their
equivalence, and the analytic continuation of the solutions.  In
section \ref{sec:Coulomb}, we present the 2D Coulomb gas
representation of the self-similar solution and, finally, in section
\ref{sec:SLC}  we investigate different limits and special cases
which are of particular interest for physics.

\section{Schwarz Function}\label{sec:SF}

The Schwarz function \cite{Davis} plays a very important role in
what follows, so we devote some space here to introduce this
object. Let a curve ${\cal C}$ in the plane be given by an
equation
\begin{equation}
F(x,y)=0,
\end{equation}
where $x$ and $y$ are cartesian coordinates in the plane. Replacing
them by complex coordinates $z=x+iy$, $\bar{z}=x-iy$, we rewrite the
equation for the curve in the form
\begin{equation}
G(z,\bar{z})=F\left(\frac{z+\bar{z}}{2},\frac{z-\bar{z}}{2i}\right)=0
\end{equation}
Solving this equation with respect to $\bar{z}$ (always possible
locally for analytic curves) we find
\begin{equation}\label{eq:Schwarz}
\bar{z}=S(z)
\end{equation}
The so defined function $S(z)$ is called the Schwarz function of the
curve ${\cal C}$.  The equation (\ref{eq:Schwarz}) is complex and
thus contains two real equations, which are mutually dependent,
since each of them bears the same information as the Eq. (1).  The
constraint on $S$ is clear after taking the conjugate of equation
(\ref{eq:Schwarz}):
\begin{equation}\label{eq:unitarity}
z=\bar{S}(S(z)).
\end{equation}
This property of the Schwarz function is called unitarity \cite{MWZ}.
The Schwarz function is holomorphic and has a great advantage over a
conformal map description of curves, as it does not imply any
particular parametrization on the curve.

The geometrical meaning of the Schwarz function becomes clear if one
imagines that the curve ${\cal C}$ is a mirror.  Then for the object
at a point $z$ close enough to the mirror the image is at the
point $\overline{S(z)}$.  For instance, the Schwarz function
for a circle centered at the
origin is $S(z)=R^2/z$, where $R$ is the radius
of the circle.

Given the Schwarz function $T(w)$ of a closed curve ${\cal T}$ in
the (``mathematical") plane $w$ and the conformal map $f(w)$ from
the outside/inside of that curve onto the outside/inside of a curve
${\cal C}$ on the (``physical") plane $z$, one can easily find the
Schwarz function of the curve ${\cal C}$:
$S(z)=\bar{f}(T(f^{-1}(z)))$, where $f^{-1}$ denotes the inverse
function, and we adopt the standard notation $\bar f(z)= \overline{f
(\bar{z})}$. In particular, if the curve ${\cal T}$ in the
mathematical plane is the unit circle, we find that the Schwarz
function is given by
\begin{equation}\label{eq:Scircle}
S(z)=\bar{f}(1/w), \quad \quad z=f(w)
\end{equation}

\section{Laplacian Growth and the Schwarz function}\label{sec:LGSF}

In this section we formulate the Laplacian Growth in terms of the
Schwarz function of the boundary following \cite{Howison91}.

For a quasistatic motion of a viscous fluid between two closely
spaced parallel surfaces, the Navier-Stokes equation is reduced
to the Darcy's Law \cite{Lamb} which states that the 2D velocity
$\bf{v}$ of the fluid is proportional to the gradient of pressure
$p$.  In properly rescaled variables it has the form
\begin{eqnarray}
&&{\bf v}=-{\bf \nabla} p\label{eq:darcy}\\
&&\Delta p =0 \label{eq:continuity}
\end{eqnarray}
where the last equation (\ref{eq:continuity}) follows from the
incompressibility condition $\mbox{div}\, \bf{v}=0$. In the air
(zero viscosity) domain pressure is constant, $p=\mbox{const}$.
Neglecting surface tension, one concludes that $p(x,y)$ is a
harmonic function in the oil domain constant along the boundary. In
this paper we consider only a single sink placed at infinity. In
this case pressure $p$ behaves at infinity as
\begin{equation}
p(x,y) \simeq -\frac{Q}{2\pi}\log R, \quad R\to \infty ,
\end{equation}
where $R^2=x^2+y^2$ and $Q$ is the pumping rate (that is the
decrease of the oil area per unit time).

Introducing the holomorphic complex potential $W(z)=-p+i\psi$ and
the complex velocity $v=v_x+iv_y$, one can rewrite equation
(\ref{eq:darcy}) in the form $v=\overline{\partial_z W}$. In the
case of a single air bubble in the cell the complex potential is
given by
\begin{equation}\label{eq:Ww}
W(z,t) = \frac{Q}{2\pi}\log w(z,t)
\end{equation}
Here $w(z,t)$ is (time dependent) conformal map from the oil domain
onto the exterior of the unit circle normalized in such a way that
$w(z, t)\simeq z/r(t)$ as $|z|\to \infty$ with a real positive
coefficient $r(t)$. This coefficient is called the (external)
conformal radius
of the air domain. In what follows we set $Q=\pi$. In these units,
area of the growing domain is equal to $\pi t$.

Let us describe the oil/air interface by its time-dependent Schwarz
function $S(z,t)$.  If at time $t$ a point $z$ is at the boundary,
$\bar{z}=S(z)$, then at time $t+dt$ the point $z+vdt$ belongs to the
new boundary, i.e., $\overline{z+vdt}=S(z+vdt,t+dt)$. Therefore, on
the boundary it holds
\begin{equation}\label{eq:infinitezimal}
\bar{v}=\dot{S}+S'v,
\end{equation}
where dot and prime denote partial derivatives with respect to $t$ and $z$.

The unit tangent vector to the boundary equals to $dz/dl$, where $l$
is the arclength $dl =\sqrt{dzd\bar{z}}=dz \sqrt{S'}$, so
\begin{equation}\label{eq:tangent}
dz/dl =\frac{1}{\sqrt{S'}}
\end{equation}

In the case of zero surface tension pressure is constant along the boundary,
and so its gradient is orthogonal to it.
In the complex notation we have
\begin{equation}
\mbox{Re}\frac{\bar{v}}{\sqrt{S'}}=0.
\end{equation}
From unitarity (\ref{eq:unitarity}) we find $S'=1/\bar{S'}$.
Together with the previous equation this gives
\begin{equation}
\bar{v}+vS'=0.
\end{equation}
Using equation (\ref{eq:infinitezimal}) and (\ref{eq:Ww}) (with
$Q=\pi$) we then find
\begin{equation}\label{eq:LGS}
\dot{S}=2\bar{v}=2\partial_z W = \partial_z \log w
\end{equation}
After analytical continuation from the boundary, this equation
is valid everywhere in the oil domain.
Equation (\ref{eq:LGS}) connects the change on the Schwarz function
of the boundary with the solution to the Dirichlet
boundary value problem for the
oil domain with given sources/sinks.

One can represent the Schwarz function in the form
\begin{equation}\label{eq:S+-}
S(z)=S_{-}(z)+S_{+}(z)
\end{equation}
where $S_{-}$ is regular in the oil domain, and $S_{+}$
is regular in the air domain. The function
$S_{+}(z)$ is given by the Cauchy integral
\beq\label{basic3}
S_{+}(z,t)=\frac{1}{2\pi i} \oint_{\p \DD (t)}
\frac{\bar z' dz'}{z' - z}
\eeq
where $\DD$ is the air domain and the point $z$ is
assumed to be inside it. (A similar Cauchy integral, with the point
$z$ outside $\DD$, can be written for $S_{-}$.)
As the function $\partial_z W$ has no singularities in the oil domain, we find
\begin{equation}\label{eq:S+dot}
\dot{S}_{+}(z)=0
\end{equation}
From this we conclude that all harmonic moments of the oil domain
\begin{equation}\label{eq:harmMoments}
t_k=
\frac{1}{2\pi i k}\oint_{\p \DD} \bar z z^{-k}dz
\frac{1}{2\pi i k}\oint_{\p \DD} S_{+}(z)z^{-k}dz, \quad
k\geq 1,
\end{equation}
are integrals of motion.  This fact was observed by S.Richardson
\cite{Richardson}.

\section{Self-similar solutions. General relations}\label{sec:SSSGR}

Self-similarity expresses the fact that the evolution of the domain
is just a dilatation of its shape.  It means that if initially the
Schwarz function of the boundary is $S(z)$, then at any other time
$\bar{z}/r(t)=S(z/r(t))$ for the points on the boundary, where the
real dilatation parameter, $r(t)$, initially equals to $1$. In other
words,
\beq\label{SS2} S(z,t)=r(t)S(z/r(t)).
\eeq
Let us take $r(t)$
to be the conformal radius of the growing (air) domain. It is defined as
the derivative at infinity of the conformal mapping from the
exterior of the unit circle onto the oil domain normalized by the
conditions that infinity is sent to infinity and $r(t)>0$. Then
$S(z)$ is the Schwarz function of the oil domain scaled so that its
conformal radius is $1$.

The self-similar reduction of eq.(\ref{eq:LGS}) reads
\beq
\label{SS3} S(z)-zS'(z)=2A \, \p_z \log w(z)
\eeq
where $r^{2}(t)= t/A$ and $w(z)$ is the conformal map of the domain
with conformal radius $1$. Taking the contour integral of equation
(\ref{SS3}) over the boundary we find that the constant $A$ is the
area (divided by $\pi$) of the domain with conformal radius $1$.

Since the r.h.s. of (\ref{SS3}) is an analytic function in the
exterior of $\DD$, this equation implies the relation
$$
S_{+}(z)-zS_{+}'(z)=0
$$
with the general solution \beq\label{SS4} S_{+}(z)=sz \eeq where $s$
is a constant. This solution corresponds to an ellipse with the axes
ratio $(1+s)/(1-s)$. We thus see that {\it regular} self-similar
solutions are exhausted by ellipses.

This result is not surprising and can be seen right from equations
(\ref{eq:S+dot}) and (\ref{eq:harmMoments}). Under the dilatations
$z\rightarrow rz$ the $k$th harmonic moment changes
as $t_{k}\rightarrow t_{k}r^{2-k}$. On the other hand, the harmonic
moments are integrals of motion and as such cannot change. This
leaves as with the only possibility to have all the moments equal to
$0$ except $t_{2}$, which is exactly the statement (\ref{SS4}).

\begin{figure}
\includegraphics[width=2in,height=2in]{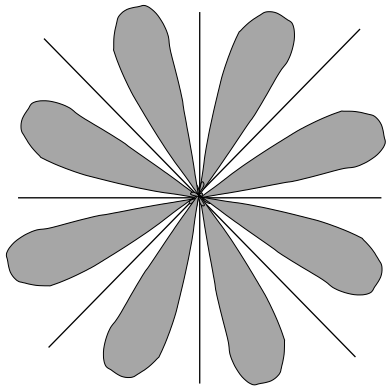}
\caption{\label{fig:SelfCrossing} A singular self-crossing domain.
The thick lines are the branch cuts of the function $S_{+}$.
Spacings between two adjacent fingers are called fjords.}
\end{figure}

However, there is a family of {\it singular} self-similar solutions
of equation (\ref{SS3}) for which the interface may cross itself at
the origin, see Fig. \ref{fig:SelfCrossing} (to be prepared
for the next section, the domain
looks like $\ZZ _N$-symmetric one
but in general no additional symmetry is implied).
For such solutions the
constant $s$ in (\ref{SS4}) is different in different sectors of the
plane and, therefore, $S_{+}$ has jumps across some rays with
$\mbox{arg}\, z = \mbox{const}$ starting at $0$ and coming to
$\infty$.  We stress that these rays are singularities of $S_{+}$,
so they do not depend on $t$. Let $\mit\Gamma$ be such a ray. Take
the jump of the both sides of eq. (\ref{SS3}) on this ray. Since the
r.h.s. is regular, we get
$$
[S(z)]_{\mit\Gamma}-z[S'(z)]_{\mit\Gamma}=0\,,
\quad z\in \mit\Gamma
$$
(the jump is defined as $[S(z)]_{\mit\Gamma}=
S(ze^{+i0})-S(ze^{-i0})$, $z\in \mit\Gamma$), hence
$$
[S(z)]_{\mit\Gamma}=c(\mit\Gamma )\, z\,,
\quad z\in \mit\Gamma
$$

\begin{figure}
\includegraphics[width=2in,height=2in]{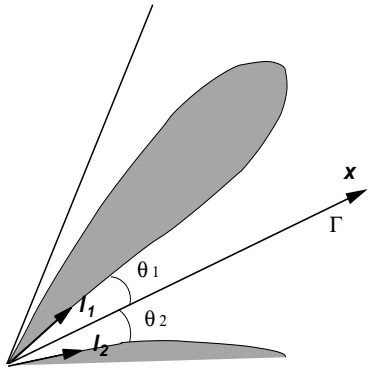}
\caption{\label{fig:jump}
The jump of the Schwarz function is related to the fjord angle.}
\end{figure}

The jump of $S(z)$ on the branch cut is related to the geometry of
the interface around the origin.  Let us choose coordinates so that
the branch cut is along the positive real semi-axis. Consider a
vicinity of the point $z=0$. The function $S'(z)$ has a constant
jump on the positive part of the real axis equal to $c=c(\mit\Gamma )$.
On the other hand, we know that the unit tangent vector to the
boundary curve is (\ref{eq:tangent}) $l (z)=1/\sqrt{S'(z)}$, so
using the notation in Fig. \ref{fig:jump}, we write
$l_{1}^{-2}=e^{-2i\theta_1}= S'(0_{+})$,
$l_{2}^{-2}=e^{-2i(\pi -\theta_2 )}= S'(0_{-})$ (here $0_{\pm}$
means $\epsilon e^{\pm i0}$ as $\epsilon \to 0$) and so the jump is
$$
c=S'(0_+)-S'(0_-)=e^{-2i\theta_1}-e^{2i\theta_2}
$$
Introducing $\theta = \theta_1 + \theta_2$ (the
fjord \footnote{Spacings between two adjacent fingers are called
fjords.} angle) and $\phi = \theta_1 -\theta_2$, we obtain the
desired relation \beq\label{SS5} c=-2ie^{-i\phi}\sin \, \theta \eeq
We see that if $c$ is purely imaginary, then the fjord is symmetric
($\phi =0$) with respect to the real axis.

The connection between the jump of $S_{+}$ and the fjord angle can
be understood purely geometrically. Let us consider a symmetric
fjord, i.e., the ray $\mit\Gamma $ is the symmetry axis.  Very close to
the origin the boundaries are almost straight lines.  Let us
consider a point $ze^{+i0}$, where $z$ is on $\mit\Gamma$. Its
reflection in the boundary lies at $ze^{i\theta}=\overline{S(ze^{+i0})}$.
For the point $ze^{-i0}$ we have
$ze^{-i\theta}=\overline{S(ze^{-i0})}$, so
$$
[S(z)]=S(ze^{+i0})-S(ze^{-i0})=(e^{-i\theta}-e^{i\theta })z=cz,
$$
from which equation (\ref{SS5}) follows.

Using the representation of the Schwarz function via conformal map
(\ref{eq:Scircle}), equation (\ref{SS3}) can be rewritten in the
familiar form \beq\label{separation} \bar f (w^{-1}) \p_w f(w)-
 f(w) \p_w \bar f (w^{-1}) =\frac{2A}{w}.
\eeq
The constant $A$ can now be expressed as
\beq\label{A}
A=\frac{1}{2\pi i} \oint_{|w|=1} \bar f (w^{-1}) d f(w).
\eeq

We note that equation (\ref{separation}) looks as the
Wronskian of $f(w)$ and $\bar f(w^{-1})$. This fact suggests that
they are two linearly independent solutions of a second order
differential equation of the form
$$
w^2 f''(w)+wf'(w)+V(w)f(w)=0
$$
with a real ``potential" $V$, such that $V(w)=\bar V(w^{-1})$. Below
we shall see that this is indeed the case.

\section{Symmetric self-similar solutions and
solutions in the wedge}\label{sec:W}

In this section, we study a family of self-similar solutions with the
$\ZZ _N$-symmetry (Fig. \ref{fig:SelfCrossing}).
Equivalently, these are self-similar solutions in
the wedge with the angle $\theta_0 = 2\pi/N$.

Let $\omega =e^{2\pi i/N}$ be the primitive root
of unity of degree $N$. The $\ZZ _N$-symmetry means that
if $z \in \DD$, then $\omega z \in \DD$.
For the conformal map this mean that
$f(\omega w)=\omega f(w)$ and for the Schwarz function
that $S(\omega z)=\omega^{-1}S(z)$. Moreover, from the
integral representation (\ref{basic3}) we see that the
latter property holds for $S_{+}$ and $S_{-}$ separately.

\begin{figure}
\includegraphics[width=2in]{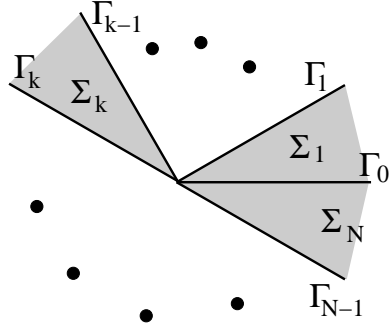}
\caption{\label{fig:SectorsAndCuts}
Thick lines $\Gamma_{k}$ are the branch cuts in $S_{+}$.
Enumeration of sectors
$\Sigma_{k}$ is also shown.}
\end{figure}

We need to introduce the following notation: Let $\mit\Sigma_k$ be
the sector $\frac{2\pi}{N}(k-1) \leq \mbox{arg}\, z \leq
\frac{2\pi}{N}k$ and let $\mit\Gamma_k = \mit\Sigma_{k+1} \cap
\mit\Sigma_k$ be the ray $\mbox{arg}\, z = \frac{2\pi}{N}k$, see
Fig. \ref{fig:SectorsAndCuts}.

\subsection{The integral equation}

Combining self-similarity ($S_{+}(z)
=s_k z$ if $z\in \mit\Sigma_k$) with $\ZZ _{N}$-symmetry
($S_{+}(\omega z)=\omega^{-1} S_{+}(z)$) we find that
$s_{k+1}=\omega^{-2}s_k$ and so
\beq\label{SZ1}
s_k =\omega^{-2k}s_0
\eeq
Therefore,
\beq\label{SZ2}
[S(z)]_{\mit\Gamma_k} =c\omega^{-2k} z\,,
\quad z\in \mit\Gamma_k
\eeq
where we have set
$c=(\omega^{-2}-1)s_0$. These jumps are singularities
of $S_{+}$\footnote{One can
restore $S_{+}$ from its jumps by means of the
Cauchy type integral:
$$
S_{+}(z)=\frac{1}{2\pi i} \sum_{k=0}^{N-1}
\int_{\mit\Gamma_k} \frac{[S_{+}(z')]_{\mit\Gamma_k}
dz'}{z'-z}
=\frac{c}{2\pi i}\int_{0}^{\infty} \sum_{k=0}^{N-1}
\frac{xdx}{\omega^k x-z}
=\,\, -\, \frac{c}{2\pi i}\sum_{k=0}^{N-1}
\omega^{2k} z \, \log(-\omega^{k}z)
$$
that is equal to $\omega^{-2k}s_0z$ in the sector
$\mit\Sigma_k$ as it must be.}.

The definition of the Schwarz function written in the form
\beq\label{SZ3} \bar f(w)=S(f(w^{-1})) \eeq implies that
singularities of $\bar f(w)$ inside the unit circle exactly
correspond to singularities of $S(z)$ outside the domain $\DD$,
i.e., to singularities of $S_{+}(z)$. In our case $\bar f (w)=f(w)$
due to the reflection symmetry with respect to the real axis, and
singularities of $f(w)$ are jumps on $N$ intervals $\mbox{arg}\, w
=\frac{2\pi}{N}k$, $0<|w|<1$ inside the unit circle, which we denote
by $\gamma_k$. Therefore, we can write $f(w)$ as the Cauchy type
integral \beq\label{SZ4} f(w)=w+\frac{1}{2\pi i} \sum_{k=0}^{N-1}
\int_{0}^{\omega^k} \frac{[f(v)]_{\gamma_k}dv}{v-w} \eeq (the
integral is over $\gamma_k$). From (\ref{SZ3}) we conclude that
\beq\label{SZ5}
\left. \phantom{A/B} [f(w)]_{\gamma_k}=-\,
[S_{+}(z)]_{\mit\Gamma_{N-k}} \right |_{z=f(w^{-1})}=
-c\omega^{-2k}f(w^{-1})
\eeq
(The minus sign is due
to the interchange of the two sides of the branch cut when one
inverts $w$ \footnote{It comes from the fact that $1/(x+i0)=1/x-i0$,
so the inversion interchanges the branches.}.)
Plugging this into (\ref{SZ4}), one
obtains, after some transformations:
$$
f(w)=w -\frac{c}{2\pi i}\sum_{k=0}^{N-1}
\omega^{-2k} \int_{1}^{\infty} \frac{f(\omega^k x)\, dx}{(1-
\omega^k xw)x}
$$
Since $f(\omega^k x)=\omega^k f(x)$,
the sum is $\sum_k \omega^{-k}/(1-\omega^k xw)$ which is
easily calculated to be
$Nxw/(1-x^N w^N)$, and so we obtain the integral equation
for the conformal map:
\beq\label{SZ6}
f(w)=w -\frac{cN}{2\pi i}\int_{1}^{\infty}
\frac{wf(x)\, dx}{1-w^N x^N}
\eeq
with the ``boundary condition" $f(1)=0$.
Let us denote the fjord angle by $\theta = \pi \beta$, then
$c=-2i \sin \, \pi \beta$ (from eq. (\ref{SS5})) and
the integral equation acquires the form
\beq\label{integral}
f(w) =w\, + \,  \frac{N\sin \, \pi \beta}{\pi}
\int_{1}^{\infty} \frac{ wf(v)\, dv}{1-w^N v^N}\,,
\quad \quad f(1)=0
\eeq
Without using eq. (\ref{SS5}), one can find
the coefficient in front of the integral by noting
that the fjord angle $\pi \beta$ implies that
$f(w)\simeq (w-1)^{\beta}$ near the point $w=1$.
Expanding eq. (\ref{SZ6}) around this point, one
obtains $c=-2i \sin \, \pi \beta$.

\subsection{Geometrical interpretation}\label{sec:geom}

There is another way to obtain the Equation (\ref{SZ6}), namely,
from the geometrical arguments applied to multi-logarithmic
solutions of 2D LG \cite{MS}.

We start from a simple observation that each logarithm in the
standard multi-logarithmic solution (anzats) for the conformal map
of the LG process asymptotically describes a single ``U'' shaped
fjord, a fjord with parallel walls. Let us consider such an anzats:
\begin{equation}\label{eq:anz}
f(w)=rw+\sum_{k}\alpha_{k}\log \left(1-a_{k}/w \right)
\end{equation}
where $\alpha_k$'s are constants of motion, while all $a_{k}$
($|a_{k}|<1$) and $r$ are functions of time. Asymptotically at large
time $r\rightarrow \infty$ and $|a_{k}|\rightarrow 1$. Integrals of
motion for such a solution are
\begin{equation}\label{eq:beta}
\beta_{k}=f(1/\bar{a}_{k})=r/\bar{a}_{k}+
\sum_{k'}\alpha_{k'}\log \left(1-a_{k'}\bar{a}_{k} \right)
\end{equation}
At large time, when all $a_{k}$s are close to $1$, we can neglect all
logarithms except one. Imagine for simplicity that $a_{0}$ is real
and $a_{0}\rightarrow 1$, then we can write
\begin{eqnarray}
&&\beta_{0}\approx r/a_{0}+\alpha_{0}\log (1-a_{0}^{2})\approx
r+\alpha_{0}\log (1-a_{0})+\alpha_{0}\log 2,      \label{eq:beta0}\\
&&f(1)\approx r+\alpha_{0}\log (1-a_{0})=\beta_{0}-\alpha_{0}\log 2.
\label{eq:f1}
\end{eqnarray}
This means that asymptotically the bottom of the fjord is at the point
$\beta_{0}-\alpha_{0}\log 2$.

Let us now take a number $\delta $ such that $\delta \ll 1$ and
$\delta \gg 1-a_{0}$, then using $e^{i\delta }\approx 1+i\delta $
we can calculate
\begin{eqnarray}
&&f(1+i\delta )\approx r+\alpha_{0}\log (i\delta ),\nonumber\\
&&f(1+i\delta )-f(1-i\delta ) = i\pi \alpha_{0}. \nonumber
\end{eqnarray}
This shows that the width of the fjord is constant and
equals $\pi |\alpha_{0}|$. The direction of the fjord is given by
the phase of $\alpha_{0}$.

Using this simple correspondence between the logarithmic
singularities and the ``U'' shape fjords we can construct a
so-called ``V'' shape fjord, i.e., a fjord which walls form a
constant angle between them.

It is clear that for this construction we need to put many identical
fjords in such a way that they have equal distance between their
vertices).  All fjord vertices must be on a ray which direction is
the same as the direction of the fjord. For the $\ZZ _{N}$-symmetric
case we then write:
\begin{eqnarray}
&&f(w)=rw+\sum_{l=0}^{\infty }\sum_{k=0}^{N-1}
\alpha \omega^{k}\log (1-a_{l}\omega^{k}/w), \label{eq:fDiscret}\\
&&\beta_{l}=l\Delta =f(1/a_{l}),\label{eq:betaDiscret}
\end{eqnarray}
where $a_{l}$, $\alpha $, $\Delta $ are real and $\Delta $ is the distance
between the nearest $\beta $s. The numbering of $a_{l}$ starts at the largest
$a$, which is $1$, so
\begin{equation}\label{eq:l0}
a_{0}=1.
\end{equation}
We now want to take the limit of $\Delta, \alpha \rightarrow 0 $. At this
limit the spacing between $a$'s becomes small, and we can think of $a_{l}$
as a smooth function $a(l)$, or alternatively $l(a)$. We then can go from
summation to integration and write
\begin{eqnarray}
&&f(w)=rw+\alpha \sum_{k=0}^{N-1}
\omega^{k}\int_{0}^{\infty } \log (1-a\omega^{k}/w)dl
=rw+\alpha \sum_{k=0}^{N-1}
\omega^{k}\int_{0}^{1}\frac{l(a)}{a-w\omega^{-k}}da,
\label{eq:fCont}\\
&&l(a)=\frac{1}{\Delta }f(1/a),\label{eq:betaCont}
\end{eqnarray}
where we integrate by parts and use (\ref{eq:l0}).
Substituting eq. (\ref{eq:betaCont}) into (\ref{eq:fCont}), we get
\begin{equation}\label{eq:fGeom}
f(w)=rw+\frac{\alpha }{\Delta }\sum_{k=0}^{N-1}\omega^{k}\int_{0}^{1}
\frac{f(1/a)}{a-w\omega^{-k}}
\end{equation}
from which (\ref{SZ6}) follows.
We note that the relation $\beta_{l}=l\Delta $ (see (\ref{eq:betaDiscret}))
is crucial for the derivation and it is this relation that leads to the
linear integral equation.

\section{Integral and differential equations for
the conformal map}\label{sec:IndDiff}

\subsection{The integral equation. Direct solution}

In the previous section we have derived the integral equation for the
conformal map $f(w)$ (see (\ref{integral})). It is an inhomogeneous
equation. Let us transform it into a homogeneous one. Putting $w=1$
in (\ref{integral}), we get
$$
\frac{N\sin \, \pi \beta}{\pi}
\int_{1}^{\infty} \frac{ f(v)\, dv}{ v^N -1}\, =\, 1
$$
that allows us to rewrite (\ref{integral}) as
$$
f(w)=\frac{N\sin \, \pi \beta}{\pi}
\int_{1}^{\infty}\frac{v^N (w^N -1)w}{(v^N -1)
(w^N v^N -1)} \, f(v) \, dv.
$$
After the substitution
$$
f(w)= w(1-w^{-N})^{\beta} F(w^{-N})\,,
\quad w^{-N}=x\,, \quad \kappa =2/N
$$
the equation acquires the form \beq\label{DS2} (1-x)^{\beta -1} F(x)
= \frac{\sin \, \pi \beta}{\pi} \int_{0}^{1}
\frac{t^{-\kappa}(1-t)^{\beta -1}}{1-xt} \, F(t) \, dt \eeq which
makes sense for arbitrary $\kappa <1$. The structure of the r.h.s.
suggests to look for solutions in hypergeometric functions.
Specifically, we use the identity
$$
\int_{0}^{1}\frac{t^{c-1}(1-t)^{a-1}}{1-xt}
\phantom{a}_{2}F_{1} \left (
\begin{array}{c}a, \,\, b\\c \end{array}
; t \right ) dt
$$
$$
=\, (1-x)^{-1} \, \frac{\Gamma (a) \Gamma (c-b)}{\Gamma (c-b+a)}
\phantom{a}_{3}F_{2} \left (
\begin{array}{c}a, \,\,\, c\! -\! b, \,\,\, 1 \\
c, \, c \! - \! b\! + \! a
\end{array}; \frac{x}{x-1} \right )
$$
(taken from \cite{GR}, section 7.512).
Notice that at $c-b+a=1$ the
function $_{3}F_{2}$ reduces to
$_{2}F_{1}$ and the r.h.s. becomes
$$
(1-x)^{-1} \, \Gamma (a) \Gamma (1-a)
\phantom{a}_{2}F_{1} \left (
\begin{array}{c}a, \, 1\! -\! a\\c
\end{array}; \frac{x}{x-1} \right ).
$$
Using the identities
$$
\Gamma (a) \Gamma (1-a)=\frac{\pi}{\sin \, \pi a}
$$
and
$$
\phantom{a}_{2}F_{1} \left (
\begin{array}{c}a, \, b\\c
\end{array}; z \right )
=(1-z)^{-a}
\phantom{a}_{2}F_{1} \left (
\begin{array}{c}a, \,\, c\! -\! b\\c
\end{array}; \frac{z}{z-1} \right )
$$
we see that the function
$$
F(x)=
\phantom{a}_{2}F_{1} \left (
\begin{array}{c}\beta , \,\, \beta \! -\! \kappa \\
1-\kappa \end{array}; x \right )
$$
solves eq. (\ref{DS2}).

Summing up, we have obtained the explicit
formula for the conformal map:
\beq\label{confmap}
f(w)=w(1-w^{-2/\kappa})^{\beta}
\phantom{a}_{2}F_{1} \left (
\begin{array}{c}\beta , \,\, \beta \! -\! \kappa \\
1-\kappa \end{array}; w^{-2/\kappa} \right ).
\eeq
For $\ZZ _N$-symmetric patterns $\kappa =2/N$
and $\pi \beta$ is the fjord angle. Equivalently,
this solution describes the self-similar growth
in the wedge with angle $\theta_0 =2\pi/N$.
Being restricted to the sector $0\leq \mbox{arg}\, w
\leq \pi \kappa$, the r.h.s. of (\ref{confmap})
makes sense for all $0< \kappa <1$ and provides
the solution for the self-similar growth in the wedge
with angle $\theta_0 =\pi \kappa$.

We note some simple properties of the map (\ref{confmap}). First,
there is the symmetry $\beta \rightarrow 1-\beta $. Second, at $\beta =0$
the hypergeometric function is identically $1$ and the function
(\ref{confmap}) is simply $f(w)=w$ that corresponds to the self-similar
growth of a disk or sector. In the
opposite limit, the function (\ref{confmap})
gives the conformal map to the complement of a needle-like
domain (a degenerate finger).
Let us consider two cases separately: i) $0<\kappa <1/2$ and
ii) $1/2<\kappa <1$ (the case $\kappa >1$
is discussed in Appendix A). In case i), at $\beta =\kappa$
the hypergeometric function is again identically $1$ and
$f(w)=(w^{1/\kappa }-w^{-1/\kappa })^{\kappa }$
is the map to the complement of a needle growing from the origin.
In case ii), at $\beta =1-\kappa $ we use the symmetry
$\beta \rightarrow 1-\beta $ and again get the same mapping.
Note that in case ii)
for values of $\beta$ in the range $1-\kappa < \beta < 1/2$ a critical
point of the function (\ref{confmap}) (a value of $w$ at which $f'(w)=0$)
is outside the unit circle, so the map (\ref{confmap}) is no longer
conformal. Such values of $\beta$ are thus non-physical.

To summarize, the physically meaningful range of the parameters
is as follows.
If $\kappa$ does not exceed $\frac{1}{2}$, then $\beta$ can be any
number between $0$ and $\kappa$, and $\pi (\kappa - \beta )$ is the
angle of the finger (the angle between the tangent lines to the
finger boundary at the origin). If $\frac{1}{2} < \kappa < 1$, then
$\beta$ can not exceed $1-\kappa$.
As $\beta$ approaches $1-\kappa$, the finger becomes a very thin
needle. Formally, in this limit the angle of the finger tends to
$\pi (\kappa - \beta )$ (not to zero!)
but the neighborhood of the origin where
this angle properly describes the form of the finger vanishes
as $\beta$ tends to $1-\kappa$.

With the conformal map (\ref{confmap}) known explicitly, one is able
to find some geometric characteristics of the growing self-similar
fingers. In particular, the distance from the finger tip to the
origin is
$$
R= |f(e^{i\pi \kappa /2)}| =2^{\beta}\!\!
\phantom{a}_{2}F_{1} \left (
\begin{array}{c}\beta , \,\, \beta \! -\! \kappa \\
1-\kappa \end{array}; -1\right )
$$
(the conformal radius is $1$). Using some identities for the
hypergeometric function, one can represent it in a more explicit
form:
\beq\label{DS3}
R=2^{\beta -1} \frac{\Gamma \left (
\frac{\beta -\kappa}{2} \right ) \Gamma (1-\kappa )}{\Gamma (\beta -
\kappa ) \Gamma \left (1- \frac{\beta +\kappa}{2} \right )}.
\eeq
The area of the finger with conformal radius $1$ will be found
below. The result is $\frac{1}{2}\pi \kappa A$, where the constant
$A$ (see (\ref{A})) is given by
\beq\label{A2}
A=\frac{\Gamma
(1-\kappa )\Gamma (-\kappa)}{\Gamma (\beta - \kappa ) \Gamma
(1-\beta -\kappa )}.
\eeq
For physical values of the parameters $\kappa$ and $\beta$
this constant is always positive.
For non-physical values ($1/2<\kappa <1$ and
$1-\kappa <\beta <1/2$) formula (\ref{A2}) gives a negative
area.

\subsection{Differential equations for the conformal map}

It turns out that the
solution to the integral equation (\ref{integral})
obeys a differential equation of second order
(that is not surprising because the solution is expressed
through the hypergeometric function).
For the first time it was written in \cite{Tu}.
Specifically, the equation reads
\beq\label{differ1}
w\p_w (w \p_w f)+ N^2\, \frac{\beta
(1-\beta )w^N}{(w^N -1)^2}\, f = f.
\eeq
It has two linearly independent solutions:
\beq\label{f}
f(w)=w(1-w^{-N})^{\beta}
\phantom{a}_{2}F_{1} \left (
\begin{array}{c}\beta , \,\, \beta \! -\! \frac{2}{N} \\
1-\frac{2}{N} \end{array}; w^{-N} \right ),
\eeq
\beq\label{g}
g(w)=w^{-1}(1-w^{-N})^{\beta}
\phantom{a}_{2}F_{1} \left (
\begin{array}{c}\beta , \,\, \beta \! +\! \frac{2}{N} \\
1+\frac{2}{N} \end{array}; w^{-N} \right ).
\eeq
The solution $f(w)$ (growing at infinity as $w$)
is our conformal map (\ref{confmap}).
The meaning of the second solution (tending to $0$
at infinity) is clarified below.

It is useful to present the integral and differential equations in
another form which may be more suggestive for some purposes. Set
$w=e^{2x/N}=e^{\kappa x}$ and $f(e^{\kappa x})=\psi (x)$. In this
notation, the equations are:
\begin{itemize}
\item
The integral equation:
\beq\label{integral2}
\psi (x)=e^{\kappa x}+\frac{2\sin \, \pi \beta}{\pi}
\int_{0}^{\infty} \frac{e^{\kappa (x+x')}
\psi (x')dx'}{1-e^{2(x+x')}},
\eeq
\item
The differential equation:
\beq\label{differ2}
\psi '' (x) +\frac{\beta (1-\beta )}{\mbox{sinh}^2 x}
\, \psi (x) =\kappa ^2 \psi (x).
\eeq
\end{itemize}
The latter is the Schroedinger equation in the
singular potential $V(x)=-\,
\frac{\beta (1-\beta )}{\mbox{sinh}^2 x}$. We need
solutions on the positive real axis such that $\psi (0)=0$.
The equation has two linearly independent solutions
$\psi_{\pm \kappa}(x)$, where
$$
\psi_{\kappa}(x)=e^{\kappa x}
(1-e^{-2x})^{\beta}
\phantom{a}_{2}F_{1} \left (
\begin{array}{c}\beta , \,\, \beta \! -\! \kappa \\
1-\kappa \end{array}; e^{-2x} \right ).
$$
The solution that grows at infinity, $\psi_{+\kappa}(x)$,
corresponds to the conformal map.

In order to show that $\psi_{\kappa}$ obeys both
integral and differential equations, we rewrite them
as the eigenvalue equations
$\hat I \psi = \frac{\pi}{\sin \, \pi \beta}\psi$,
$\hat D \psi = \beta (\beta -1)\psi$ for the operators
$$
\hat I \psi (x)=\int_{0}^{\infty}
\frac{e^{\kappa (x+x')}\, \mbox{sinh}\, x}{\mbox{sinh}\,
(x+x')\, \mbox{sinh}\, x'}\, \psi(x')\, dx' ,
$$
$$
\hat D \psi (x)=\mbox{sinh}^2 x (\p_{x}^{2}-\kappa^2)
\psi (x).
$$
A direct calculation shows that the operators $\hat I$, $\hat D$
commute on the space of functions vanishing at the origin, and thus
have a family of common eigenfunctions. The correspondence between
the eigenvalues is established by their local behavior around the
origin.

A change of the variables leads to another form
of the differential equation, where the parameters
$\kappa$ and $\frac{1}{2}-\beta$ get interchanged.
Namely, introduce the new variable $\xi$ such that
\beq\label{xi}
\mbox{cosh}\, \xi = \mbox{coth}\, x
\eeq
(equivalent forms: $\mbox{sinh}\, \xi =-(\mbox{sinh}\, x)^{-1}$,
$\xi = \log \, \mbox{tanh}\, (x/2)$) and set
$$
\tilde \psi (\xi )=(\mbox{sinh} \, \xi)^{1/2}
\psi (x(\xi )).
$$
Then eq. (\ref{differ2}) becomes
\beq\label{differ3}
\tilde \psi '' -
\frac{\kappa^2 -\frac{1}{4}}{\mbox{sinh}^2 \xi}
\, \tilde \psi
=\left (\beta -\frac{1}{2}\right )^2 \tilde \psi.
\eeq

\subsection{Analytic continuation of the conformal map}

An important information is encoded in the analytic
continuation of the function $f(w)$ to the interior
of the unit circle. It can be achieved by means of the
identity
$$
\phantom{a}_{2}F_{1} \left (
\begin{array}{c}a , \, b \\
c \end{array}; z \right )
=\frac{\Gamma (c)\Gamma (b-a)}{\Gamma (b)\Gamma (c-a)}
\, (-z)^{-a}
\phantom{a}_{2}F_{1} \left (
\begin{array}{c}a , \,\, a-c+1 \\
a-b+1 \end{array}; z^{-1} \right )
$$
$$
\phantom{aaaaa}
+\,\,\, \frac{\Gamma (c)\Gamma (a-b)}{\Gamma (a)\Gamma (c-b)}
\, (-z)^{-b}
\phantom{a}_{2}F_{1} \left (
\begin{array}{c}b , \,\, b-c+1 \\
b-a+1 \end{array}; z^{-1} \right ).
$$
In our case this identity gives:
$$
f(w)\, =\,
A w(1-w^{-N})^{\beta} (-w^{-N})^{-\beta}
\phantom{a}_{2}F_{1} \left (
\begin{array}{c}\beta , \,\, \beta +\frac{2}{N} \\
1+\frac{2}{N} \end{array}; w^N \right )
$$
$$
\phantom{aaaaa}
+\,\, \frac{\sin \, \pi \beta}{\sin \frac{2\pi}{N}}
\,  w(1-w^{-N})^{\beta} (-w^{-N})^{-\beta +\frac{2}{N}}
\phantom{a}_{2}F_{1} \left (
\begin{array}{c}\beta , \,\, \beta -\frac{2}{N} \\
1-\frac{2}{N} \end{array}; w^N \right ),
$$
where \beq\label{A1} A=\frac{\Gamma (1-\kappa )\Gamma (-\kappa
)}{\Gamma (\beta - \kappa ) \Gamma (1-\beta -\kappa )}\,, \quad
\kappa = 2/N . \eeq Taking appropriate branches of the multi-valued
functions, we have
$$
(1-w^{-N})^{\beta} (-w^{-N})^{-\beta}=
(1-w^N)^{\beta},
$$
$$
(-w^{-N})^{2/N}= \omega^{2k-1}w^{-2} \quad \mbox{if
$w\in \mit\Sigma_k$}.
$$
Therefore, we can compactly write the analytically
continued function in the form
\beq\label{cont}
f(w)=B_k f(w^{-1})+
Ag(w^{-1})\,,
\quad w\in \mit\Sigma_k \,,
\eeq
where
\beq\label{Bk}
B_k =\omega^{2k-1}\,
\frac{\sin \, \pi \beta}{\sin \frac{2\pi}{N}}\,,
\eeq
$A$ is given by (\ref{A1}),
and $g(w)$ is the second solution (\ref{g}) of the
differential equation (\ref{differ1}).
(Note that $f(w^{-1})$ and $g(w^{-1})$ are also
solutions of this equation.)

In fact (\ref{cont})
is nothing else than the decomposition of the
Schwarz function $S=S_{+}+S_{-}$. To see this,
we rewrite (\ref{cont}) in the equivalent form
$$
f(w^{-1})=B_{1-k} f(w)+
Ag(w)\,,
\quad w\in \mit\Sigma_k \,,
$$
which makes it clear (\ref{eq:Scircle}) that the l.h.s. is
$S(z)$ while the first term in the r.h.s. is
$S_{+}(z)$ (cf. (\ref{SZ2}), (\ref{SZ5})).
Therefore, the second term is just $S_{-}(z)=g(f^{-1}(z))$.
This observation allows us to compute the area
of the domain with conformal radius $1$ and to
conclude that it is equal to $\pi A$, in agreement
with the previously introduced notation (\ref{A}).
Indeed, we write:
$$
\mbox {Area}\, (\DD ) =\frac{1}{2i}\oint_{\p \DD}
S(z)dz =\frac{1}{2i}\oint_{\p \DD}
S_{+}(z)dz + \frac{1}{2i}\oint_{\p \DD}
S_{-}(z)dz\,.
$$
The first integral vanishes since $S_+$ is analytic
inside $\DD$ while the second one is
$$
\frac{A}{2i}\oint g(w) df(w) =\pi A
$$
(by taking the residue at infinity).

\section{The 2D Coulomb gas representation
for the self-similar solutions}\label{sec:Coulomb}

As it has been shown in \cite{MWZ,KKMWZ,TBAZW}, the LG
dynamics (with zero surface tension) can be simulated by the 2D
Coulomb gas in an external field in the limit ${\cal N}\to \infty$,
where ${\cal N}$ is the number of charges:
\beq\label{C1}
Z_{{\cal N}} (t) = \int \prod_{i<j}^{{\cal N}} |z_i - z_j |^2
\prod_{l=1}^{{\cal N}} e^{\frac{{\cal N}}{t}W(z_l , \bar z_l )} d^2
z_l \,,
\eeq
where the potential $W$ is determined by the ``positive"
part of the Schwarz function:
\beq\label{C2} W(z, \bar z)=-|z|^2 + 2
{\cal R}e \, \int_{0}^{z}S_{+}(z')\, dz' \,.
\eeq
Specifically, it can
be shown that in the limit ${\cal N}\to \infty$ the charges
uniformly fill a domain $\DD (t)$ in the plane and the shape of this
domain depends on $t$ according to the Laplacian growth equation.
For the self-similar $\ZZ _{N}$-symmetric solutions a simple
calculation gives
\beq\label{C3} W(z, \bar z)=-|z|^2 + \frac{\sin \,
\pi \beta}{2\sin \, \frac{2\pi}{N}} \left ( e^{-2\pi i(2k-1)/N}z^2 +
e^{2\pi i (2k-1)/N}\bar z^2 \right ) \, , \quad z\in \mit\Sigma_k\,.
\eeq
This potential is continuous across the boundary between the
sectors and enjoys the $\ZZ _{N}$-symmetry.

The free energy of the Coulomb gas (\ref{C1}), normalized
in a special way,
\beq\label{C4}
{\cal F}(t)=t^2  \lim_{{\cal N}\to \infty}
\frac{\log Z_{{\cal N}}}{{\cal N}^2}\,,
\eeq
plays an important role. (As is shown in \cite{KKMWZ},
it is related to tau-function of certain integrable hierarchies.)
To find it for the self-similar solutions, we use
the integral representation
\beq\label{C5}
2\pi {\cal F}(t)=\int_{\DD (t)} W(z, \bar z)d^2 z +
t \int_{\DD (t)}\log |z|^2 \, d^2 z
\eeq
and the relation
\beq\label{C6}
\frac{\p^2 {\cal F}(t)}{\p \, t^2} = 2\log \, r(t)
\eeq
($r(t)$ is the external conformal radius
of the domain $\DD (t)$) obtained in \cite{KKMWZ}.
Since the potential (\ref{C3}) is homogeneous,
the self-similar ansatz plugged into (\ref{C5})
yields
$$
2\pi {\cal F}(t)=r^4 (t) \int_{\DD (A)} W\, d^2z
+Ar^4(t)\int_{\DD (A)}\log (r^2 (t)|z|^2 ) d^2 z\,,
$$
where the domain $\DD (A)$ has conformal radius $1$
and $r^2 (t)=t/A$. Therefore, we can write
$$
{\cal F}(t)\, =
\, r^4(t){\cal F}(A)+\frac{A^2}{2} r^4(t)\log r^2(t) \, =\,
\frac{t^2}{A^2}{\cal F}(A) +\frac{1}{2} t^2 \log \frac{t}{A}
$$
and calculate the second $t$-derivative:
$$
\frac{\p^2 {\cal F}}{\p \, t^2} =
\frac{2}{A^2}{\cal F}(A) + \log \frac{t}{A} +\frac{3}{2}\,.
$$
The r.h.s. must be equal to $2\log r(t) = \log(t/A)$, hence
${\cal F}(A)=-\frac{3}{4}A^2$ and eventually we obtain
\beq\label{C7}
{\cal F}(t)=\frac{1}{2}t^2 \log \frac{t}{A} \, -\, \frac{3}{4}\, t^2\,,
\eeq
where $A$ is given in (\ref{A1}).

\section{Special and limiting cases}\label{sec:SLC}

The trivial special cases are $\beta =0$ (no fjords,
the domain is the disk, $f(w)=w$) and
$\beta =2/N$ (the domain is a collection of $N$
symmetric needles growing from $0$, $f(w)=w(1-w^{-N})^{2/N})$.
A non-trivial special case, where the solution is available
in elementary functions is $N=4$ (wedge with the right
angle). At $N=4$, the conformal map is \cite{TRHC}
$$
f(w)=\frac{w}{2}\left [
(1-w^{-2})^{\beta}
(1+w^{-2})^{1-\beta} +
(1+w^{-2})^{\beta}
(1-w^{-2})^{1-\beta}
\right ]
$$
and $A=(1-2\beta )\cos  \pi \beta$.
A direct substitution $N=1$ leads to a singularity. To resolve it,
we set $N=1-\epsilon$, $\beta =4\epsilon t_2$ and tend $\epsilon \to
0$. The general formulas then give: $S_{+}(z)=2t_2 z$, $f(w)=w
+2t_2/w$, $A=1-4t_{2}^{2}$, i.e., the domain is an ellipse with the
second harmonic moment $t_2$ (in the normalization used in
\cite{MWZ}). The case $N=2$ gives, in a similar way, the same family
of ellipses (restricted to the half-plane). See also Appendix A.

The most interesting limit is $N\to \infty$. In this case
we expect the solutions to reproduce, after a proper
rescaling, the Saffman-Taylor
fingers in the channel geometry.
In the rest of this section we show
that this is indeed the case.

First of all, let us change $w^N \to w$ and consider the
conformal map
$$
f_1 (w)=w^{\kappa /2} (1-w^{-1})^{\beta}
\phantom{a}_{2}F_{1} \left (
\begin{array}{c}\beta , \,\, \beta -\kappa \\
1-\kappa \end{array}; w^{-1} \right )
$$
in the limit $\kappa = 2/N \to 0$ and $\beta \to 0$ so that
$\beta/\kappa$ remains a finite parameter.
We set
$$
\beta = \kappa (1-\lambda )\,,
$$
so that $\pi \lambda \kappa$ is the angle of the
finger (tending to zero in the limit).
Second, passing to the channel of a finite
width (say $2\pi$) requires the rescaling
of the physical plane $z \to Nz$. At last,
in order to be closer to the finger tip
one should shift the origin. Taking this into account,
it is not difficult to see that the correct
limit to the channel geometry is
\beq\label{ch1}
f_{{\rm ch}} (w)= \lim_{N\to \infty}
 \left [ Nf_1 (w) -N\right ]
\eeq Performing this limit with the $f_1(w)$ given above, one
obtains \beq\label{ch2} f_{{\rm ch}} (w) = \log w +2(1-\lambda )
\log (1-w^{-1}) \eeq which is the standard form of the
Saffman-Taylor finger with the relative width $\lambda$. Note that
the hypergeometric function behaves as $1+ O(\kappa^2)$ and so does
not contribute to $f_{{\rm ch}}(w)$.

It is also instructive to derive the limiting form
of the time-dependent conformal map and to follow in detail
how self-similarity in the wedge
turns into translational invariance in the channel.
At nonzero $\kappa$ the time evolution is just a similarity
transformation, so we can write, taking into account the shift
of the origin:
$$
\tilde f_{{\rm ch}}(w,t)=\lim _{N \to \infty}
\left (\frac{r(T_0)+t}{r(T_0)}\, N\, f_1 (w)\, -
\, N \right )
$$
where $T_0 =AN^2$, i.e., $r(T_0)=N$.
Expanding
$$
\frac{r(T_0)+t}{r(T_0)} =
\left (1+\frac{t}{T_0}\right )^{1/2}=1+ \frac{t}{2T_0} +\ldots
$$
we get
$$
\tilde f_{{\rm ch}}(w,t)=f_{{\rm ch}}(w)+
\frac{tN}{2T_0}\,.
$$
It remains to rescale time as
$t=2NT$, where $T$ is kept finite
(note that still $t\ll T_0$),
then
the last term becomes $T/A$. The limiting
value of $A$ is $1-\beta \kappa^{-1} = \lambda$,
so we obtain the time-dependent Saffman-Taylor
finger in the standard form
$$
f_{{\rm ch}}(w,T)=\tilde f_{{\rm ch}}(w,t)=
\frac{T}{\lambda}+
\log w +2(1-\lambda ) \log (1-w^{-1})\,.
$$

\acknowledgments

Ar.A. was supported by the Oppenheimer fellowship at LANL.
M.M. gratefully acknoweldges generous support
of the LDRD project 20030037DR at LANL during
completion of this work.
The work of A.Z. was supported in part by
RFBR grant 06-02-17383 and by grant NSh-8004.2006.2
for support of scientific schools.

\appendix

\section{Shapes of the domains}\label{sec:forms}

In this Appendix we discuss the
range of parameters of the conformal map (\ref{confmap})
and the corresponding
shapes of the air domains. It is convenient to adopt
the function (\ref{confmap}) to the case of a single sector in the physical
plane:
\begin{equation}\label{eq:confmapSingle}
f(w)=w^{\kappa /2}(1-1/w)^{\beta}
\phantom{a}_{2}F_{1} \left (
\begin{array}{c}\beta , \,\, \beta \! -\! \kappa \\
1-\kappa \end{array}; 1/w \right ).
\end{equation}
(we remind that in our notation the wedge angle
is $\Theta_{w}=\pi \kappa $ and the fjord angle
is $\Theta_{fjord}=\beta \pi $).

First, we notice that at $\beta =0$ the hypergeometric function is
identically $1$, so the conformal map is just $f(w)=w^{\kappa /2}$ --
the map from the exterior of the unit circle to the
exterior of the sector.

Further investigation should be done separately in the different regions:
i) $0<\kappa <1/2$, ii) $1/2<\kappa <1$, iii) $1<\kappa <3/2$,
and iv) $3/2<\kappa <2$. Two special points $\kappa =1$ and $\kappa =2$
require separate considerations.

In the region i) the allowed values of $\beta $ are $0<\beta <\kappa $.
At $\beta =\kappa $ the hypergeometric function is again identically $1$.
The map then is $f(w)=(w^{1/2}-w^{-1/2})^{\kappa }$ and
the air domain is a needle. The conformal map has a critical point
(a point where $f'(w)=0$) $w=-1$ right on the unit circle.
If we formally consider
the case $\beta >\kappa $ then we see that the critical point
is outside the
unit circle. The map (\ref{eq:confmapSingle}) is then no longer conformal.

In the region ii) the allowed values of $\beta $ are $0<\beta <1-\kappa $.
At $\beta =1-\kappa $ we use the symmetry
$\beta \rightarrow 1-\beta $ to see
that the map is again given by
$f(w)=(w^{1/2}-w^{-1/2})^{\kappa }$ and
the air domain is again a needle. The
critical point is at $w=-1$. If we increase the value of $\beta$, then the
critical point moves outside the unit circle and the map
is no longer conformal.

In the region iii) the allowed values of $\beta $ are $0<\beta <\kappa-1$.
At $\beta =\kappa-1$ the hypergeometric function is just $1+1/w$.
The function (\ref{eq:confmapSingle}) is then
$f(w)=(w^{1/2}-w^{-1/2})^{\kappa -1}(w^{1/2}+w^{-1/2})$. This gives
a needle-like domain but the needle is
``two-sided" and rotated by $\pi /2$ (see the figure).
The map has two
critical points on the unit circle at the
points $w=\exp (\pm i\phi_{0})$, where
$\tan \phi_{0}=\sqrt{\kappa -1}$. This corresponds to the two tips of the
needle. If we increase the value of $\beta$, then both
critical points move outside the unit circle and the map
is no longer conformal.

In the region iv) the allowed values of $\beta $ are $0<\beta <2-\kappa$.
At $\beta =2-\kappa $ we use the symmetry
$\beta \rightarrow 1-\beta $ and again see that the
situation is identical to the one in region iii).
There are two critical points right on the unit
circle. Both of them go outside should we increase the
value of $\beta $ even further.

All the four cases are illustrated in the figure.

\begin{tabular}{cccc}
\begin{minipage}{0.2\columnwidth}
\includegraphics[width=\columnwidth]{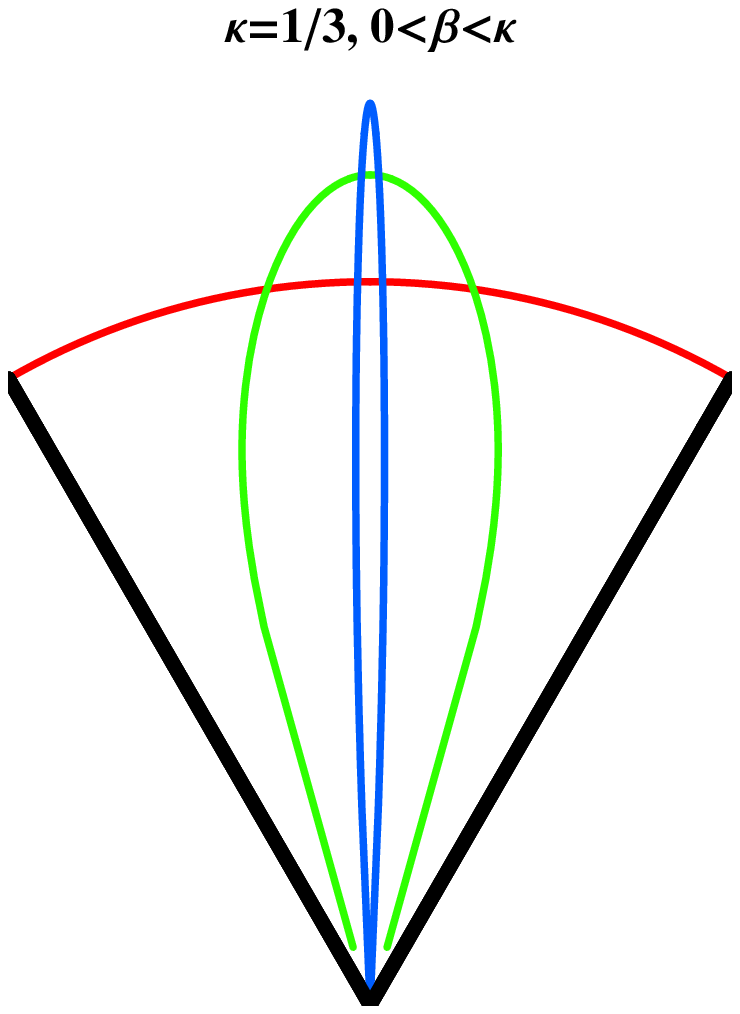}
\end{minipage}
&
\begin{minipage}{0.22\columnwidth}
\includegraphics[width=\columnwidth]{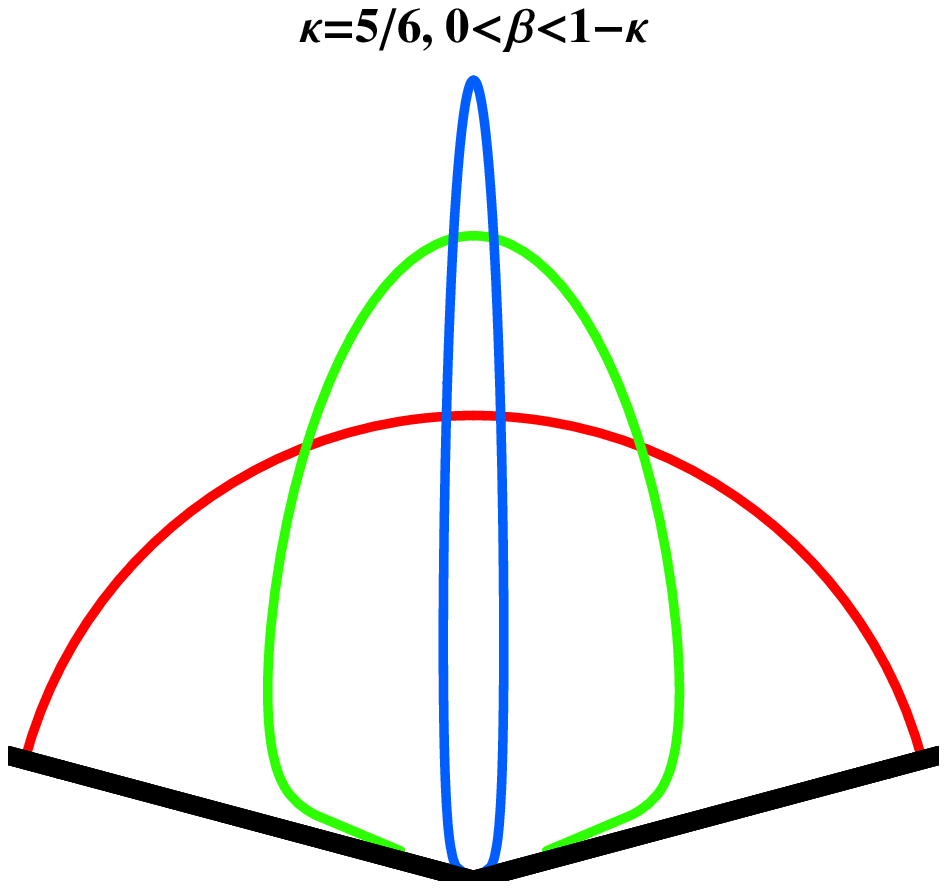}
\end{minipage}
&
\begin{minipage}{0.28\columnwidth}
\includegraphics[width=\columnwidth]{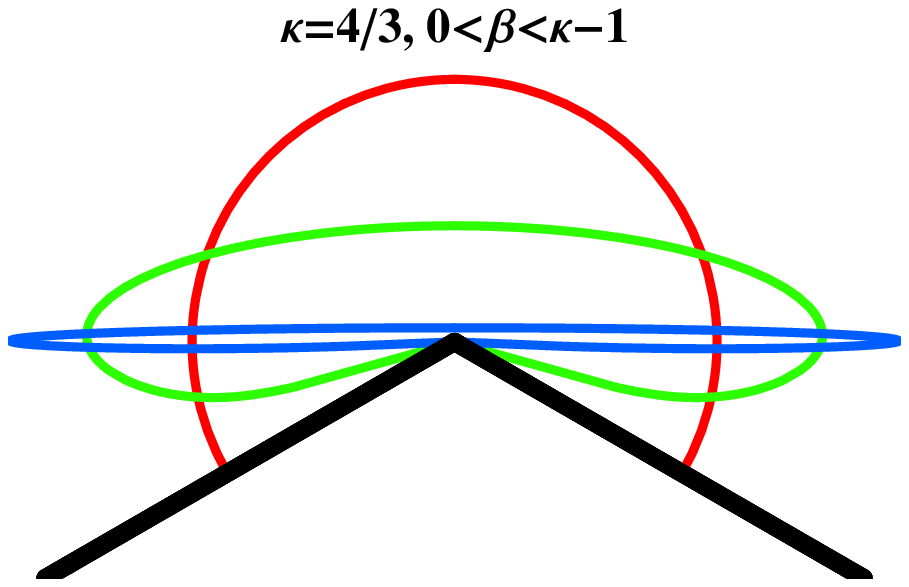}
\end{minipage}
&
\begin{minipage}{0.3\columnwidth}
\includegraphics[width=\columnwidth]{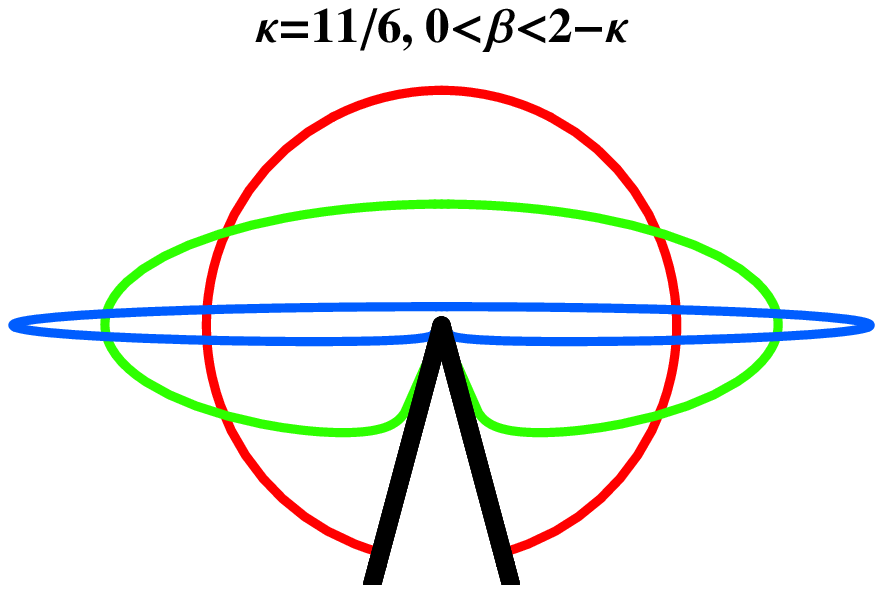}
\end{minipage}
\end{tabular}
\vspace{.1in}

The special value $\kappa =1$ requires careful consideration. We set
$\kappa =1-\delta $ and take the limit $\delta  \rightarrow 0$,
$\beta \rightarrow 0$ with the arbitrary ratio $a=\beta /\delta <1$.
The hypergeometric function is then $1-a/w$ and the conformal map
(\ref{eq:confmapSingle}) becomes $f(w)=w^{1/2}-aw^{-1/2}$ which gives
half of an ellipse with axes $1-a$ and $1+a$.

At the special  value $\kappa =2$,  we set
$\kappa =2-\delta $ and again take the limit $\delta  \rightarrow 0$,
$\beta \rightarrow 0$ with an arbitrary ratio $a=\beta /\delta <1$.
The hypergeometric function becomes $1-a/w^{2}$ and the conformal map
$f(w)=w-a/w$ gives an ellipse with axes $1-a$ and $1+a$.

\end{document}